\documentclass[letter,twocolumn]{jpsj3}
\usepackage{color}
\usepackage{bm}
\usepackage{txfonts}
\usepackage[resetlabels,labeled]{multibib}

\title{Low-temperature Magnetic Fluctuations Investigated by $^{125}$Te-NMR on the Uranium-based Superconductor UTe$_{2}$}

\author{Hiroki Fujibayashi$^1$\thanks{fujibayashi.hiroki.w49@kyoto-u.ac.jp}, Katsuki Kinjo$^{1}$\thanks{Present adress: Institute of Multidisciplinary Research for Advanced Materials, Tohoku University, Sendai, Miyagi 980-8577, Japan}, Genki Nakamine$^{1}$, Shunsaku Kitagawa$^{1}$, Kenji Ishida$^1$\thanks{kishida@scphys.kyoto-u.ac.jp}, \\Yo Tokunaga$^{2}$, Hironori Sakai$^{2}$, Shinsaku Kambe$^{2}$, Ai Nakamura$^{3}$, Yusei Shimizu$^{3}$, \\Yoshiya Homma$^{3}$, Dexin Li$^{3}$, Fuminori Honda$^{3}$, and Dai Aoki$^{3,4}$}
\inst{$^1$Department of Physics, Kyoto University, Kyoto 606-8502, Japan \\
$^2$Advanced Science Research Center, Japan Atomic Energy Agency, Tokai, Ibaraki 319-1195, Japan \\
$^3$IMR, Tohoku University, Oarai, Ibaraki 311-1313, Japan \\
$^4$University Grenoble, CEA, IRIG-PHERIQS, F-38000 Grenoble, France} 

\abst{To investigate the static and dynamic magnetic properties on the uranium-based superconductor UTe$_{2}$, we measured the NMR Knight shift $K$ and the nuclear spin-lattice relaxation rate $1/T_{1}$ in $H \parallel a$ by $^{125}$Te-NMR on a $^{125}$Te-enriched single-crystal sample.  
$1/T_1T$ in $H \parallel a$ is much smaller than $1/T_1T$ in $H \parallel b$ and $c$, and magnetic fluctuations along each axis are derived from the $1/T_1T$ measured in $H$ parallel to all three crystalline axes.
The magnetic fluctuations are almost identical at two Te sites and isotropic at high temperatures, but become anisotropic below 40 K, where heavy-fermion state is formed.
The character of magnetic fluctuations in UTe$_2$ is discussed with the comparison to its static susceptibility and the results on other U-based superconductors.
It is considered that the magnetic fluctuations probed with the NMR measurements are determined by the magnetic properties inside the two-leg ladder formed by U atoms, which are dominated by the $q_a$ = 0 ferromagnetic fluctuations.}

\begin{document}
\maketitle

\begin{figure}[tbp]
\begin{center}
\includegraphics[width=8cm]{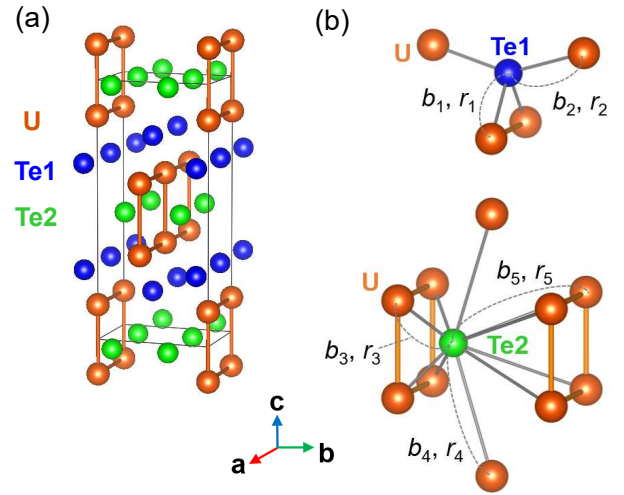}
\end{center}
\caption{(Color online) Crystal structure of UTe$_2$ emphasizing the two-leg ladder structure formed by the U atoms. Two crystallographically inequivalent Te sites, $4j$ and $4h$, with the point symmetries $mm2$ and $m2m$ are shown as Te1 and Te2, respectively\cite{VESTA}. (b) Atomic arrangement around each Te site. $b_j$s and $r_j$s indicate the transfer-field parameters and the distances between U and Te site, respectively. Here, $r_1=3.0553~\AA$, $r_2=3.1817~\AA$, $r_3=3.1648~\AA$, $r_4=5.2706~\AA$, and $r_5=5.3462~\AA$\cite{HutanuACSB2020}.}
\label{f1}
\end{figure}

Superconductivity in UTe$_{2}$ [Superconducting (SC) transition temperature $T_{\mathrm{SC}}$ = 1.6 K] was discovered at the end of 2018\cite{RanScience2019}, and was confirmed by other group immediately\cite{AokiJPSJ2019}.  
UTe$_{2}$ has the orthorhombic crystal structure with the space group $Immm$ (\#71, $D^{25}_{2h}$), which belongs to the symmorphic group [Fig.\ref{f1}(a)]\cite{VESTA}. 
UTe$_{2}$ is considered to be a spin-triplet superconductor because its characteristic SC properties, such as extremely high SC upper critical fields $H_{\rm c2}$\cite{RanScience2019,RanNatueP2019} and the magnetic-field ($H$) enhancement of $H_{\rm c2}$\cite{RanScience2019}, are quite similar to those observed in uranium (U)-based ferromagnetic (FM) superconductors\cite{AokiJPSJRev2019} (UGe$_2$ under pressure\cite{SaxenaNature2000}, and URhGe\cite{AokiNature2001} and UCoGe\cite{HuyPRL2007} at ambient pressure), which are spin-triplet superconductors.  
However, different from these FM superconductors, UTe$_{2}$ shows superconductivity in the paramagnetic state, and does not show any FM ordering even in the high-pressure region where superconductivity suddenly disappears\cite{BraithwaiteComnPhys2019, ThomasSciAd2020}.
More surprisingly, inelastic neutron scattering (INS) experiments have revealed the predominance of the incommensurate antiferromagnetic (AFM) spin fluctuations with the wave-vector $\bm{Q}_{\rm ic}$ = (0, 0.57, 0) \cite{DuanPRL2020,knafoPRB2021}.
Since the U atoms in UTe$_2$ form a two-leg ladder with legs along the $a$ axis and rungs along the $c$ axis [Fig. \ref{f1}(a)], the AFM interaction is parallel to the inter two-leg ladder direction along the $b$ axis. 
Thus, one of the great interests in UTe$_{2}$ is to understand the magnetic properties in the normal state, particularly whether FM fluctuations exist or not inside the two-leg ladder structure above $T_{\rm SC}$, although the INS measurements could not detect any FM fluctuations\cite{DuanPRL2020,knafoPRB2021}. 
To clarify this issue, $^{125}$Te-NMR measurements for $H \parallel a$ on a single-crystal UTe$_2$ are crucially important, because the directional magnetic fluctuations can be known as discussed later. 

In the previous measurement\cite{TokunagaJPSJ2019}, we could not observe $^{125}$Te-NMR signals in $\mu_0 H \sim 2$ T below 20 K for $H \parallel a$ due to the divergence of the nuclear spin-spin relaxation rate $1/T_{2}$ and the broadening of the spectrum.
To overcome this difficulty, we prepared a $^{125}$Te-enriched single crystal sample with a nearly cubic shape, and successfully observed the $^{125}$Te-NMR signals in low fields parallel to the $a$ axis. 
We measured the NMR Knight shift $K$ and nuclear spin-lattice relaxation rate $1/T_{1}$ under magnetic field in $H \parallel a$ down to 1.5 K. 
Together with previous NMR results\cite{TokunagaJPSJ2019}, magnetic fluctuations along each crystalline axis are derived.
It was found that magnetic fluctuations are almost isotropic at high temperatures, and becomes highly anisotropic below 40 K, as in the case of the static susceptibility. 

Single-crystal UTe$_{2}$ was grown by the chemical-vapor-transport method with iodine as the transport agent\cite{RanScience2019,AokiJPSJ2019}. 
Natural U and 99.9\% $^{125}$Te-enriched metals were used as starting materials for the present sample. 
The $^{125}$Te (nuclear spin $I=1/2$, gyromagnetic ratio $^{125}\gamma_{n}/2\pi=13.454$ MHz/T)-NMR measurements were performed on a single crystal of $0.8\times1.2\times0.9$ mm$^{3}$ size. 
The frequency-sweep NMR spectrum was obtained using the Fourier transform of a spin-echo signal observed after the spin-echo radio-frequency pulse sequence with a 5-kHz step in a fixed magnetic field. 
The magnetic field was calibrated using $^{63}$Cu ($I$ = 3/2, $^{63}\gamma/2\pi=11.285$ MHz/T) and $^{65}$Cu ($I$ = 3/2, $^{65}\gamma/2\pi=12.089$ MHz/T) NMR signals arising from the NMR coil. 
We used the split SC magnet, which generates a horizontal field, and combined it with a single-axis rotator to apply a magnetic field in each crystalline direction precisely.

\begin{figure}[tbp]
\begin{center}
\includegraphics[width=8cm]{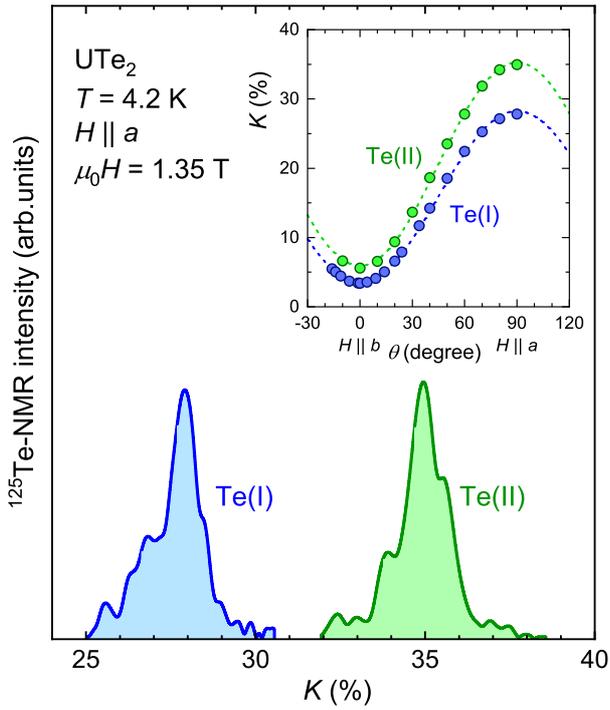}
\end{center}
\caption{(Color online) $^{125}$Te-NMR spectrum in $H \parallel a$ shown against the Knight shift of $K = (f - f_0)/f_0 $. The smaller [larger]-$K$ peak is denoted as a Te(I) [Te(II)] peak. The inset shows the magnetic-field angle dependence of the Te(I) and Te(II) peaks in the $ab$ plane. }
\label{f2}
\end{figure}
Figure \ref{f2} shows the $^{125}$Te-NMR spectra for $H \parallel a$, which are shown against the NMR shift $K = (f - f_0)/f_0 $. 
Here, $f$ is the NMR frequency, and $f_0$ is the reference frequency determined as $f_0 = (^{125}\gamma_n/2\pi) \mu_0 H$.     
There are two crystallographically inequivalent Te sites, $4j$ and $4h$, with the point symmetries $mm2$ and $m2m$ in UTe$_2$, and these are denoted as Te1 and Te2 sites [Fig.~\ref{f1}(a)], respectively. 
Correspondingly, we observed two $^{125}$Te-NMR peaks as reported in the previous paper\cite{NakamineJPSJ2019,NakaminePRB2021,NakamineJPSJ2021}.
An NMR peak with the smaller [larger] $K$ in $H \parallel b$ is assigned as Te(I) [Te(II)] by following the previous paper\cite{TokunagaJPSJ2019}.
The inset of Fig.~\ref{f2} shows the $H$ angular dependence of the NMR Knight shift at both the Te(I) and Te(II) peaks measured in the $ab$ plane at 4.2 K. 
These angle dependence are well fitted by the theoretical curves of $K(\theta) = K_a \sin^2\theta +K_b \cos^2\theta$, where $K_a$ ($K_b$) is the Knight shift in $H \parallel a$ ($b$).  
The $a$-axis aligned NMR spectra of the Te(I) and Te(II) peaks were detected by following the angle dependence in the $ab$ plane from the observed $b$-axis NMR spectra\cite{TokunagaJPSJ2019,NakamineJPSJ2019,NakaminePRB2021,NakamineJPSJ2021}. 
$K_a$ of the Te(I) and Te(II) peaks at 4.2 K is 27.9\% and 35.2\%, respectively. 
In the present measurement, the smaller magnetic field of $\mu_0 H$ = 1.35 T than in the previous study\cite{TokunagaJPSJ2019} makes the NMR peak sharper. 

From the present measurement, it turned out that the previous NMR-peak assignment in $H \parallel a$ should be reinterpreted.
Although the Te(I) and Te(II) peaks were overlapped in $H \parallel c$ in the previous study\cite{TokunagaJPSJ2019}, there are found to be no crossing nor overlapping in the angle dependence of the two peaks in the present study. 
The present peak assignment results in nearly isotropic values of the hyperfine coupling constants at the two Te sites, as shown in Table I.  
Note that the peak assignment presented here [Te(I) and Te(II)] does not correspond to the crystallographical site (Te1 and Te2).

\begin{table}[t]
\caption {Hyperfine coupling constants $A_{\rm hf}^{\alpha}$ evaluated from the linear relation between Knight shift and bulk susceptibility along $\alpha = a, b$ and $c$. 
This corresponds to the diagonal term in the hyperfine tensor.   }
\begin{tabular}{c|ccc} \hline
  & $H \parallel a$ & $H \parallel b$ & $H \parallel c$ \\
  & $A_{\rm hf}^{a}$ (T / $\mu_{\rm B}$)  & $A_{\rm hf}^{b}$  (T / $\mu_{\rm B}$)  & $A_{\rm hf}^{c}$ (T / $\mu_{\rm B}$) \\  \hline\hline
Te(I) & 3.8 & 3.4 & 3.9 \\
Te(II) & 4.7 & 5.2 & 3.9 \\ \hline
\end{tabular}
\label{table}
\end{table}

\begin{figure}[tbp]
\begin{center}
\includegraphics[width=8cm]{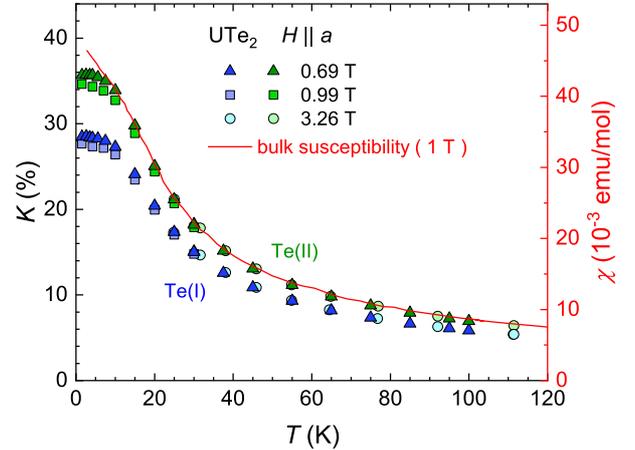}
\end{center}
\caption{(Color online) Temperature dependence of the Knight shift determined from the Te(I) and Te(II) peaks. The $K_a$ above 30 K\cite{TokunagaJPSJ2019} and $\chi_a$ above 5 K\cite{LiJPSJ2021} are quoted from the references.  }
\label{f3}
\end{figure}

Figure \ref{f3} shows the temperature ($T$) dependence of $K_a$ measured in 0.69 and 0.99 T below 100 K.
In the figure, the $K_a$ data measured in 3.26 T above 30 K\cite{TokunagaJPSJ2019} and the $\chi_a$ measured in 1 T are also shown\cite{LiJPSJ2021}. 
As reported in the paper\cite{TokunagaJPSJ2022}, the
$K_a$ is proportional to $\chi_a$ above 10 K, but is nearly saturated and remains almost constant at low temperatures. 
This is in contrast to the continuous increase in $\chi_a$ below 10 K, which would be ascribed to the extrinsic effect induced by the U-atom deficiency\cite{Haga2022JPCM}.  

\begin{figure*}[bt]
\begin{center}
\includegraphics[width=15cm]{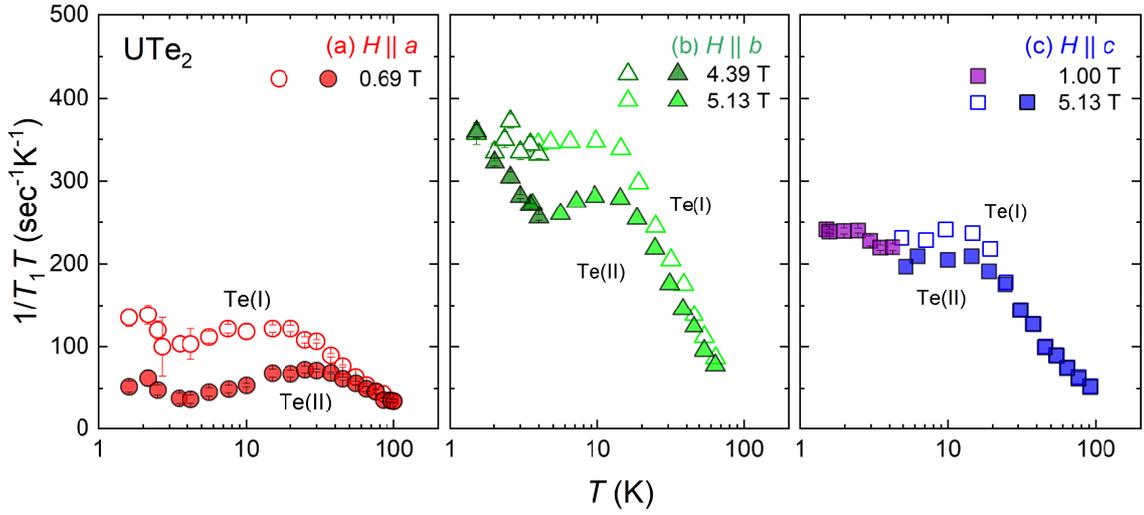}
\end{center}
\caption{(Color online) Temperature dependence of $1/T_{1}T$ at Te(I) and Te(II) in (a) $H \parallel a$, (b) $H \parallel b$ and (c) $H \parallel c$.
The data of $1/T_1T$ in $H \parallel b$ and $H \parallel c$ above 4.2 K are quoted from the reference\cite{TokunagaJPSJ2019}.
$1/T_1T$ in $H \parallel c$ below 4.2 K was measured at the broad NMR peak arising from the Te(I) and Te(II) signals. }
\label{f4}
\end{figure*}

Figures \ref{f4}(a)-(c) show the $T$ dependence of $1/T_{1}T$ of Te(I) and Te(II) measured in $H$ parallel to  each crystalline axis, respectively. 
$1/T_1T$ in all directions of Te(I) and Te(II) roughly follows the Curie-Weiss behavior with different Curie terms above 45 K, which is consistent with the resistivity behavior, suggestive of the localized $5f$ state.
With decreasing $T$, all $1/T_1T$s deviate from the Curie-Weiss behavior and almost saturate below 15 K in $H \parallel b$ and $c$.
This temperature might be related to a broad Schottky-like anomaly observed at $\sim 12$ K with the various thermodynamic, transport probes\cite{WillaPRB2021} and the nuclear spin-spin relaxation rate $1/T_2$\cite{TokunagaJPSJ2022}.
In contrast, $1/T_1T$ in $H \parallel a$ shows a broad maximum at around 20 K for Te(I) and 30 K for Te(II), respectively.
The magnitude of $1/T_1T$ at low $T$ is different between Te(I) and Te(II) in $H \parallel a$ and $b$, but the difference is small in $H \parallel c$.   
Below 4 K, 1/$T_1T$ increases again. 
This increase is stronger at Te(II), and would be relevant to the low-$T$ increase reported in the relaxation rate in the $\mu$SR\cite{SundarPRB2019} and $1/T_2$\cite{TokunagaJPSJ2022}. 

Following the previous procedure\cite{TokunagaJPSJ2019}, we derived the directional dynamic susceptibility components $R_{i,\alpha}$ for Te($i$) along each orthorhombic crystal axis $\alpha$ with using the relation of $(1/T_1T)_{i, \gamma} = R_{i, \alpha} + R_{i, \beta}$.
Here, $i$ = I and II, $\alpha, \beta, \gamma = \{ a, b, c \}$, and $(1/T_1T)_{i, \gamma}$ is the $1/T_1T$ of Te($i$) measured in $H \parallel \gamma$.
In general, $R_{i, \alpha}$ is expressed as    
\begin{equation}
R_{i, \alpha} = \sum_{\mbox{\boldmath$q$}, \xi} A_i^{\alpha\xi}(\mbox{\boldmath$q$})^2 \frac{\chi_{\xi}^{''}(\mbox{\boldmath $q$}, \omega_{\rm N})}{\omega_{\rm N}}.  
\end{equation}
with the $\omega_{\rm N}$ and $\chi_{\xi}^{''}(\mbox{\boldmath $q$}, \omega_{\rm N})$ are the NMR frequency and the imaginary part of the dynamic susceptibility along $\xi = \{a, b, c\}$, respectively.
$A_i^{\alpha\xi}(\mbox{\boldmath$q$})$ is a $\mbox{\boldmath$q$}$-dependent hyperfine coupling tensor at Te($i$), which is assumed to be expressed as $A_{i}^{\alpha\xi}(\mbox{\boldmath$q$})^2 \sim \left|A_i^{\alpha\xi}(0) F_i(\mbox{\boldmath$q$})\right|^2$ with $F_i(\mbox{\boldmath$q$}) =  \sum_{j} b_j e^{i\mbox{\boldmath$q$}\cdot\mbox{\boldmath$r$}_j}$ for simplicity.
Here, $F_i(\mbox{\boldmath$q$})$ is the site-dependent form factor, which consists of the transfer-field parameters $b_j$ from the $j$th-U atoms surrounding the Te($i$) as shown in Fig.~\ref{f1} (b), and this acts as the filter of spin fluctuations.
\begin{figure*}[t]
\begin{center}
\includegraphics[width=15cm]{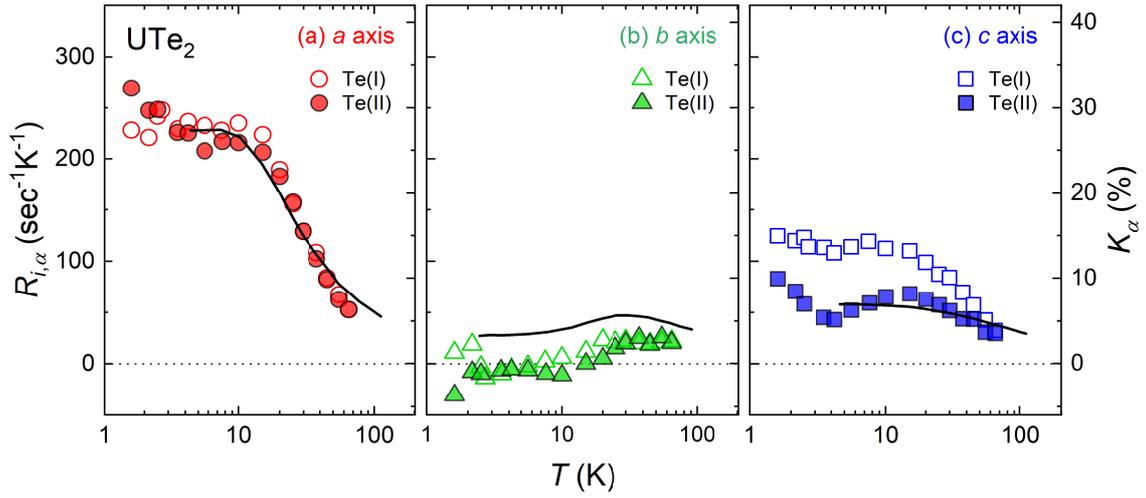}
\end{center}
\caption{(Color online) Temperature dependence of (a) $R_{i, a}$, (b) $R_{i, b}$, and (c) $R_{i, c}$ ($i$ = I and II) evaluated from the data shown in Fig.~\ref{f4}. 
Temperature dependence of $K_{\alpha}$ ($\alpha$ = $a$, $b$, and $c$) at Te(I) is also shown by the solid curves\cite{TokunagaJPSJ2019}} 
\label{f5}
\end{figure*}

Figures \ref{f5}(a)-(c) show the $T$ dependence of $R_{i, \alpha}$ derived from $1/T_1T$ of Te(I) and Te(II) signals with the above relations.
The magnitude and temperature dependence of $R_{i, \alpha}$ are almost the same between two signals, although the magnitude of $R_{c}$ at low temperatures is different.
In addition, it was found that the anisotropy of $R_{i, \alpha}$ becomes remarkable below 40 K where the $\chi$ along the $b$ axis shows a maximum. 
This temperature is called $T_{\chi_{\rm max}}$ and is regarded as the formation of the heavy-fermion state.
$1/T_1T$ results also show that the anisotropic heavy-fermion state is formed, consistent with the Knight-shift results.

Here, we consider the effect of AFM fluctuations along the $b$ axis detected by the INS measurements with $\mbox{\boldmath$Q$} = (0, q_b, 0)$\cite{DuanPRL2020,knafoPRB2021}. 
As seen in Fig.~\ref{f1}(b), the Te1 site is located at the center of the above two U atoms along the $b$ axis and the bottom two U atoms along the $a$ axis. 
The Te2 sites are at the nonsymmetric position in the unit cell, and thus, $1/T_1T$ at the Te2 site can also probe the AFM fluctuations.
The form factor at the Te$x$ ($x$ = 1 and 2) site against $q_b$ along (0, $q_b$, 0) $F_{x}(q_b)$ is calculated in the presence of the three-dimensional coupling as\cite{Kambe2010PRB},   
\begin{align*}
\left|F_1(q_b)\right|^2=&4 b_1^2 + 4b_2^2\cos^2{\left(q_b/2\right)}+8b_1b_2\cos{\left(q_b/2\right)}\\
\left|F_2(q_b)\right|^2=&16\left[|b_3 -b_5|^2+4b_3b_5\cos^2{\left(q_b/2\right)}\right]\\ 
 & +4b_4^2+16(b_3b_4+b_4b_5)\cos{\left(q_b/2\right)}.
\end{align*}
As for the form factor of the Te2 site, the second and third nearest neighbor U sites are taken into account.
As the distances between the Te1 site and the surrounding four U atoms ($r_1$ and $r_2$) are almost the same, we assumed that the $b_1$ and $b_2$ are the same ($b_1 = b_2$).
At the Te2 site, as $r_4 \sim r_5$, we assumed $b_4 = b_5$.
$b_3/b_{4 (5)}$ is the ratio of the hyperfine coupling constants for the nearest and second (third) nearest-neighbor sites at the Te2 site, and are estimated as $\sim (r_{4 (5)}/r_3)^3 \sim 4.7$ by assuming the Ruderman-Kittel-Kasuya-Yosida type interaction\cite{Kambe2010PRB}.

\begin{figure}[tbp]
\begin{center}
\includegraphics[width=7cm]{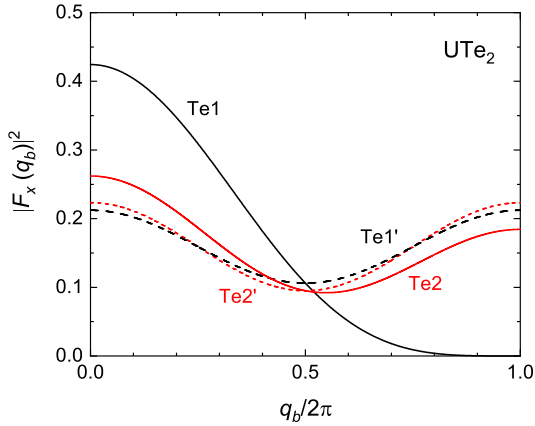}
\end{center}
\caption{(Color online) $q_b$ dependence of the hyperfine form factor $|F_{x}(q_b)|^2$ at the Te1 and Te2 sites. The dotted curves (Te1' and Te2') are the $q_b$ dependence of $|F_{x}(q_b)|^2$, which are the Te1 and Te2 form factor without the $c$-axis interladder coupling. Each form factor is normalized to be $\sum_{q_b}|F_{x}(q_b)|^2=1$.}
\label{f6}
\end{figure}
Figure \ref{f6} shows the $q_b$-dependent form factor $|F_{x}(q_b)|^2$, where each $\sum_{q_b}|F_{x}(q_b)|^2$ is normalized to the unit.
It is noted that the $q_b$ dependence of $|F_{x}(q_b)|^2$ is different between two sites, and thus, $T$ dependence of $1/T_1T$ should be different between two Te sites in general, as the site difference of $|F_{x}(q_b)|^2$ yields the difference of $\sum_{q_b} |F_{x}(q_b)|^2\chi^{''}(\mbox{\boldmath$q$}, \omega_{\rm N})$.
The difference is mainly due to the presence of the third term in $|F_1(q_b)|^2$, which arises from the $c$-axis interladder coupling.  
However, the INS measurements revealed that the interladder magnetic coupling along the $c$ axis is negligibly weak, and the AFM coupling along the $b$ axis would induce the frustration in the magnetic coupling along the $c$ axis.
Thus, it is reasonable to neglect the interladder magnetic coupling along the $c$ axis.
In this case, the terms related to the $c$-axis interladder correlation ($8b_1b_2\cos{\left(q_b/2\right)}$ and $16(b_3b_4+b_4b_5)\cos{\left(q_b/2\right)}$) become zero. Thus, the form factor at Te1 and Te2 sites are expressed as,
\begin{align*}
\left|F_{1'}(q_b)\right|^2=& 4 b_1^2 + 4b_2^2\cos^2{\left(q_b/2\right)}\\
\left|F_{2'}(q_b)\right|^2=& 16\left[|b_3 -b_5|^2+4b_3b_5\cos^2{\left(q_b/2\right)}\right]+4b_4^2.
\end{align*}
The $q_b$-dependence of the above form factors at the Te1 and Te2 sites is shown by the dotted curves (Te1' and Te2') in Fig.~\ref{f6}, which are quite similar to each other.       
Therefore, the AFM fluctuations with $q_b \sim$ 0.57 observed with the INS measurements\cite{DuanPRL2020, knafoPRB2021} are detectable in the same manner with $1/T_1T$ at both the sites.
The similar $1/T_1T$ behavior at Te(I) and Te(II) are interpreted with the absence of the interladder magnetic coupling along the $c$ axis.

To investigate the properties of the magnetic fluctuations, $R_{i,\alpha}$ is compared with $K_{\alpha}$ at Te(I), as shown in Figs.~\ref{f5}(a)-(c).  
Almost linear relations between $R_{i, a}$ and $K_{a}$ are observed.
This indicates that the diagonal component in the hyperfine coupling tensor is dominant in the $a$ axis.
It is noted that $K_a$ and $R_a$ continue to increase even below $T_{\chi_{\bf max}}$, where U-5$f$ electron is in the itinerant regime.
This implies that the most dominant contributions in $\chi_a^{''}(\mbox{\boldmath$q$}, \omega)$ is $\chi_a(0, 0)$, indicative of the dominance of the $q_a$ = 0 fluctuations.
This seems consistent with the INS result that the nearest U atoms have in-phase correlations\cite{knafoPRB2021}. 
It is suggested that the fluctuations detected with the NMR measurement would be confined in the low-energy regions ($\sim 10^{-4}$ meV), since the appreciable FM fluctuations are not detected with the INS measurements above 0.7 meV\cite{DuanPRL2020}.
We also point out that the $T$ dependence of the incommensurate AFM fluctuations is almost the same as that of $\chi(0)$\cite{knafoPRB2021}. 
This implies that the magnetic fluctuations might be $q$ independent at the low-energy region.
This indicates that the local-moment character might remain even below $T_{\chi_{\rm max}}$.
It is important to investigate the energy dependence of the magnetic fluctuations to make clear the relation between NMR and INS results. 

In contrast, $R_b$ is smaller than $K_b$ and almost zero below 10 K at Te(I) and Te(II).
This indicates that the magnetic fluctuations are absent along the $b$ direction at low $T$, although the static spin component is finite in the whole $T$ range.
Furthermore, the $R_c$ at Te(I) is significantly larger than $K_c$, although $R_c$ at Te(II) is almost the same as $K_c$ at Te(II).
This suggests the presence of the additional $R_c$, which is related to the off-diagonal component in the hyperfine coupling tensor at Te(I).
Since the Te2 site is located at the symmetric position with respect to the two-leg-ladder chain, in which the U atoms have the in-phase correlations\cite{knafoPRB2021}, it is considered that the Te2 site does not have off-diagonal components in the hyperfine coupling tensor. 
One possibility is that the average effective field along the $c$ axis can arise, when the U electronic spins near the Te1 site are aligned to the $a$ or $b$ axis with the incommensurate AFM coupling along the $b$ axis, and that the fluctuations of the $c$-axis effective field can induce the additional $R_c$. 
The difference of $R_c$ values of the Te(I) and Te(II) peaks might give useful information about the NMR-signal assignment arising from the two Te sites.
Recently, it was reported that low-$T$ $1/T_1T$ in $H \parallel b$ for the two Te sites becomes different by applying pressure ($P$), and that $1/T_1T$ of Te(I) is more enhanced than $1/T_1T$ of Te(II), which is clearly recognized near critical pressure\cite{AmbikaPRB2022}.  
Since the low-$T$ magnetic susceptibility along the $b$ axis ($\chi_b$) increases with $P$\cite{LiJPSJ2021}, the enhancement of $1/T_1T$ in $H \parallel b$ of Te(I) might be understood by the increase of $R_c$ with the above scenario.      

Finally, we compare the present results to those observed in other U-based compounds.
In the U-based heavy-fermion superconductors such as URu$_2$Si$_2$\cite{Kohara1986SSC}, UPt$_3$\cite{Lee1993PRB}, and UPd$_2$Al$_3$\cite{Kyogaku1993JPSJ}, U-5$f$ electrons show the crossover from high-$T$ localized to the low-$T$ heavy-fermion states, as observed in UTe$_2$.
In these compounds, AFM fluctuations appear to be responsible for the crossover process and dominate the low-frequency magnetic properties.
The presence of the AFM fluctuations is recognized from the comparison of the $T$ dependence between $R_\alpha$ and $K_\alpha$, and the development of $R_\alpha$ is much larger than that of $K_\alpha$ with decreasing $T$. 
On the other hand, as discussed above, $R_a$ is well scaled to $K_a$ even below $T_{\chi_{\rm max}}$ down to the lowest $T$ in UTe$_2$.
This is quite in contrast with the behavior observed in these compounds, and reminiscences the FM SC UCoGe\cite{IharaPRL2010,HattoriPRL2012} and nearly FM UCoAl\cite{Karube2012PRB}.
Our NMR results in UTe$_2$ would be dominated by the magnetic properties inside the two-leg ladder formed by the U atoms, since the two Te sites are located in the vicinity of the ladder.  

In summary, we observed the $a$-axis NMR signals down to 1.5 K, and found that there are no crossing nor overlapping of the Te(I) and Te(II) peaks in the angle dependence in the $ab$ plane. 
This makes the previous peak assignment revised: the hyperfine-coupling constants at two Te-NMR peaks are nearly isotropic.  
From the results of $H\parallel a$, the magnetic fluctuations along each axis $R_\alpha$ were derived. 
We found the predominance of the FM fluctuations along the $a$ axis, which would be the properties inside the two-leg ladder along the $a$ axis.

\begin{acknowledgments}
The authors would like to thank Y. Yanase, S. Fujimoto,  W. Knafo, K. Kaneko, G. Knebel, J.-P. Brison, S. Raymond, and J. Flouquet for their valuable discussions. 
This work was supported by the Kyoto University LTM Center, Grants-in-Aid for Scientific Research (Grant Nos. JP15H05745, JP17K14339, JP19K03726, JP16KK0106, JP19K14657, JP19H04696, JP20H00130, JP20KK0061, JP22H01168 and JP22H04933) from JSPS.
\end{acknowledgments}

\end{document}